\documentclass[%
twocolumn,
 amsmath,amssymb,
aps,
pra,
]{revtex4-1}
\usepackage{listings}
\usepackage{graphicx}% Include figure files
\usepackage{dcolumn}% Align table columns on decimal point
\usepackage{bm}% bold math
\usepackage{ulem}
\usepackage{natbib}
\usepackage[colorlinks,citecolor=red,linkcolor=blue,urlcolor=blue]{hyperref}
\usepackage{capt-of}
\usepackage{hhline}
\usepackage{float}
\usepackage{mhchem}
\usepackage[nodisplayskipstretch]{setspace}
\usepackage{calrsfs}
	\DeclareMathAlphabet{\pazocal}{OMS}{zplm}{m}{n}
\usepackage{physics}
\usepackage{appendix}
\usepackage[export]{adjustbox}

\newcommand{\bk}{\mathbf{k}}

\usepackage[dvipsnames]{xcolor}
\definecolor{pink}{rgb}{0.858, 0.188, 0.478}

\usepackage{lineno,blindtext}

\begin{document}

%\preprint{APS/123-QED}

\title{ Tuning bulk topological magnon properties with light-induced magnons}
%2. Tuning topological properties of bulk magnons with light-induced amplification\\
%3. Selective bulk topological magnon amplification via light\\
%4. Selective bulk topological magnon amplification\\
%5. Selective amplification of bulk topological magnon\\
%6. Selective amplification of bulk topological magnon via light\\
%7. Selective amplification of topological magnons\\
%8. Selective amplification of topological magnons in non-centrosymmetric magnets\\
%9. Selective amplification of topological magnons in non-centrosymmetric magnets via light} %and amplification of thermal Hall conductivity}% Force line breaks with \\
%\thanks{A footnote to the article title}%
\author{Dhiman Bhowmick\href{https://orcid.org/0000-0001-7057-1608}{\includegraphics[scale=0.12]{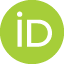}}}
 %\altaffiliation[Also at ]{Physics Department, XYZ University.}%Lines break automatically or can be forced with \\
\author{Hao Sun}
\author{Bo Yang}
\author{Pinaki Sengupta\href{https://orcid.org/0000-0003-3312-2760}{\includegraphics[scale=0.12]{orcid.png}}}%
 %\email{Second.Author@institution.edu}
\affiliation{%
 School of Physical and Mathematical Sciences, Nanyang Technological University, Singapore \\
}

\date{\today}% It is always \today, today,
             %  but any date may be explicitly specified

\begin{abstract}
%{\color{red}: Pinaki: The abstract needs to be more punchy} {\color{red} Although theoretical modelling and inelastic neutron scattering measurements have indicated the presence of topological magnon bands in multiple quantum magnets, their manifestation in experimentally measured physical observables have remained elusive. Magnons obey Bose Einstein statistics and, at equilibrium, condense at the bottom of the bands. On the other hand, the Berry curvature for most topological magnon bands is concentrated at the higher energies, and thus cannot influence equilibrium magnonic properties. In a recent work, Malz, {\it et al.}\,\cite{MagnonAmplification} have shown that topological magnons in edge states in a finite sample can be amplified using tailored electromagnetic fields. We extend their proposal to selectively generate magnons at any desired energy in isolated bands. Our results show that breaking of inversion symmetry is essential for bulk magnon amplification. Using this approach, we demonstrate the generation of bulk topological magnons in a Heisenberg ferromagnet on the breathing kagome lattice and the consequent amplification of thermal Hall effect. We discuss the application of this approach for experimental observation of thermal Hall effect in real quantum magnets.}

Although theoretical modelling and inelastic neutron scattering measurements have indicated the presence of topological magnon bands in multiple quantum magnets, experiments remain unable to detect signal of magnon thermal Hall effect in the quantum magnets, which is a consequence of magnons condensation at the bottom of the bands following Bose Einstein statistics as well as the concentration of Berry curvature at the higher energies.
In a recent work, Malz, {\it et al.}\,[Nature Communications 10, 3937 (2019)] have shown that topological magnons in edge states in a finite sample can be amplified using tailored electromagnetic fields.
We extend their approach by showing that a uniform electromagnetic field can selectively amplify magnons with finite Berry curvature by breaking inversion symmetry of a lattice.
Using this approach, we demonstrate the generation of bulk topological magnons in a Heisenberg ferromagnet on the breathing kagome lattice and the consequent amplification of thermal Hall effect.

\end{abstract}

\pacs{Valid PACS appear here}% PACS, the Physics and Astronomy
                             % Classification Scheme.
%\keywords{Suggested keywords}%Use showkeys class option if keyword
                              %display desired
\maketitle

%\tableofcontents

\section{\label{sec1}Introduction}

The successful isolation of atomically thin magnets\,\cite{Magnon4,TopologicalMagnons7,KagomeMaterial2,CrI3_P1,CrI3_P2,CrI3_P3,CrI3_P5} has triggered intensive
investigation of topological magnetic excitations in low dimensional quantum magnets.
%Following the phenomenal progress in  their fermionic counterpart,
Interest in bosonic topological phases has been rising
over the past several years since 
the band structure properties that underlie
(non-interacting) topological states are independent  of the quantum statistics of the particles. 
Topological band structures have 
%Bosonic analogues of topological phases have 
already been reported in such diverse bosonic
systems as photons~\,\cite{TopologicalPhotonicsReview1,TopologicalPhotonicsReview2,TopologicalPhotonicsReview3}, 
phonons\,\cite{TopologicalPhonons1,TopologicalPhonons2}, 
cold atoms\,\cite{TopologicalColdAtomsReview1,TopologicalColdAtomsReview2}, and 
magnons\,\cite{TopologicalMagnons1,HallEffect1,MagnonHallEffect,HallEffect3, TopologicalMagnons5,Magnon4,TopologicalMagnons7, TopologicalMagnons9}. 
Magnons, quantized low energy excitations in quantum magnets obeying Bose-Einstein statistics\footnote{The low energy excitations
can refer to either magnons, spinons or triplons 
%-- the former is a collective mode, whereas
%the latter two are localised excitations -- depending on the nature of the magnetic ground state. 
For the sake of conciseness, we refer to all magnetic excitations
generically as magnons here.}, 
are ideally suited for realizing complex bosonic phases in a controllable manner, e.g.,
Bose-Einstein condensation\,\cite{BECofMagnons}.
Microscopic modelling reveals that the time reversal symmetry-breaking Dzyaloshinskii-Moriya interaction (DMI) -- present in many quantum magnets -- imparts finite Berry curvature to non-interacting magnon bands. 
When effects of interactions are added, bosonic systems 
hold the promise of realizing new interaction driven topological phases that are not 
observed in fermionic systems, due to the different quantum statistics obeyed 
by the two. 

Magnons are charge neutral quasiparticles and hence do not exhibit one of the key signatures of fermionic topological bands, viz., the topological Hall effect where a transverse current is induced by a longitudinal potential gradient even in the absence of an external magnetic field. Instead they are expected to exhibit thermal (or magnon) Hall effect, where a longitudinal temperature gradient, $\Delta_x T$, 
 produces a current of thermally generated magnons that experiences 
a transverse force due to the geometric magnetic field, $B$,
produced by the Berry phase of the magnon bands. The resulting transverse magnon current, ${\bf J}_Q$,
constitutes a thermal Hall  effect of magnons\cite{TopologicalMagnons1,HallEffect1,MagnonHallEffect}, or magnon Hall effect (MHE), and is analogous to the Topological (or Anomalous) Hall effect in electrons.
However, while the MHE has been theoretically predicted for many quantum magnets\,\cite{TopologicalMagnons1,HallEffect1,HallEffect3,TopologicalMagnons5,TopologicalMagnons7,TopologicalMagnons9,Magnon1,Magnon3,Magnon4,Magnon5,Magnon6,Magnon7,SrCu2(BO3)2_V1,SrCu2(BO3)2_V2,WeylTriplon},
it has been  experimentally observed only in \ce{Lu2V2O7}\cite{MagnonHallEffect} and \ce{Cu[1,3-benzenedicarboxylate]}\cite{TopologicalMagnons7}. Notably, while neutron scattering experiments have shown the existence of gapped magnon bands in \ce{CrI3} and \ce{Sr2Cu(BO3)2} consistent with theoretical calculations predicting topological magnon bands, experimental efforts to observe magnon Hall effect have failed in both materials\,\cite{noTHE, noTHE2}. The reasons are threefold: (i) density of thermally excited magnons is concentrated at the band minimum  -- magnons do not obey Pauli exclusion
principle and a magnon band cannot be ``filled to the Fermi level" to observe
edge states,
(ii) the Berry curvature
in the magnetic Brillouin zone (MBZ) is often concentrated at momenta
away from the band minimum where density of thermally
excited magnons is low, and, (iii) the strength 
of intrinsic DMI in most quantum magnets is weak. These inherent
difficulties make 
observing MHE and edge states in real materials a formidable challenge. 

In a recent work, Malz, {\it et.al.}\,\cite{MagnonAmplification} have proposed that a robust edge current of magnons can be generated in a kagome ferromagnet by a {\it spatially varying} electromagnetic (EM) field. 
However, their approach, by itself, is not sufficient to amplify MHE for reasons discussed in detail later. 
In this work, we have extended their approach to excite magnons selectively at any arbitrary energy using a {\it uniform} electromagnetic field. In particular, using our approach, bulk magnons can be controllably generated in an isolated band, which is essential for amplifying MHE signals. 
Crucially, we show that breaking of inversion symmetry is necessary for selective amplification of bulk magnons and illustrate this in the breathing kagome ferromagnet.
Our results demonstrate that magnon Hall effect can be amplified by two orders of magnitude by selectively amplifying magnons at finite Berry-curvature points in reciprocal space using the proposed amplification scheme.

%both edge state and bulk magnons controllably in the experimentally important honeycomb lattice ferromagnets. Our results will be crucial in designing experiments to realize thermal Hall effect in the \ce{CrX3} family of quantum magnets.

\section{\label{sec2}Results}

\subsection{\label{sec2a} Symmetry analysis}
%{\color{red}Pinaki: Explain Malz's approach is a couple of sentences.}{\color{blue} The magnon amplification method introduced by Malz, {\it et al.}\,\cite{MagnonAmplification} uses spatially modulated electromagnetic wave which might not suitable for amplification of thermal Hall effect for a small sample size.
%We show that uniform electromagnetic can be used to selectively amplify magnons in the same band in absence of inversion symmetry.}

The amplification scheme introduced by Malz, {\it et al.}\,\cite{MagnonAmplification} relies on the use of tailored electromagnetic (EM) radiation to excite magnons. The interaction between the EM wave and the quantum magnet is described by the $ {\pazocal H}_{c} = - {\bf E}(t)\cdot \hat{\bf P}$, where ${\bf E}(t)$ is the electric field and $\hat{\bf P}$ is the polarization operator. The amplification constitutes the absorption of a photon from the incident radiation to create a pair of magnons with equal and opposite momenta, ${\pazocal H}_{c} \rightarrow \sum_{\bf k}g_{\bf k}(a_{\bf k}^\dagger a_{-\bf k}^\dagger b + h.c.)$ where $a_{\bf k}$ and $b$ represent the magnon and photon operators respectively. In their work, Malz, {\it et al.}\,\cite{MagnonAmplification} uses spatially modulated electromagnetic waves to excite edge state magnons. We argue that symmetry constraints prevent amplification of magnons by this process selectively in an isolated band in inversion symmetric lattices. Later, it will be shown that such amplification of isolated bands is crucial for amplifying physical observables such as the MHE.

%Hence the magnon amplification scheme is not applicable for amplification of thermal Hall conductivity until inversion symmetry is broken.
The Inversion symmetry operator $\hat{\pazocal{I}}$ transforms the magnon annihilation operator $\hat{\tilde{a}}_{n\bk}$ at $n$-th band at momentum $\bk$ and also transforms the electric field amplitude of light $\mathbf{E}$ as,
\begin{equation}
\hat{\pazocal{I}}\hat{\tilde{a}}_{n,\bk}=\hat{\tilde{a}}_{n,-\bk}\, ,\quad
\hat{\pazocal{I}}\boldsymbol{E}=-\boldsymbol{E}.
\end{equation}
Consequently, the magnon amplification term, which creates a pair of magnons in the same band, will transform as, 
\begin{equation}
    \hat{\pazocal{I}} \boldsymbol{E} \hat{\tilde{a}}_{n,\bk}^\dagger \hat{\tilde{a}}_{n,-\bk}^\dagger
    =
    -\boldsymbol{E} \hat{\tilde{a}}_{n,-\bk}^\dagger \hat{\tilde{a}}_{n,\bk}^\dagger
    =
    0.
\end{equation}
Thus magnons in an isolated band can not be selectively amplified in presence of inversion symmetry $\hat{\pazocal{I}}$.
%The characteristics of symmetry operator $\hat{\pazocal{O}}$ is same as inversion symmetry. %Moreover if the system posses an axis of two fold rotation along direction of propagation of light, then due to two fold rotational symmerty also.
%In the following, we 
%demonstrate this for the Heisenberg ferromagnet on a regular kagome lattice, which is the system considered by Malz, {\it et. al.}, and 
The central result of this work is the discovery that selective amplification of magnons in isolated bands can be achieved by breaking the inversion symmetry of the lattice. In the following we show this for a Heisenberg ferromagnet with additional Dzyaloshinkii-Moriya interaction (identical to the microscopic model considered in Ref.[\onlinecite{MagnonAmplification}]) on a {\it breathing} kagome lattice (that explicitly breaks the inversion symmetry).
%({\color{red}alternatively, we can say "for the system cosidered by Malz, et.al." - will that be too combative?})

\begin{widetext}

\begin{figure}[tb]
\centering
\includegraphics[width=0.95\textwidth]{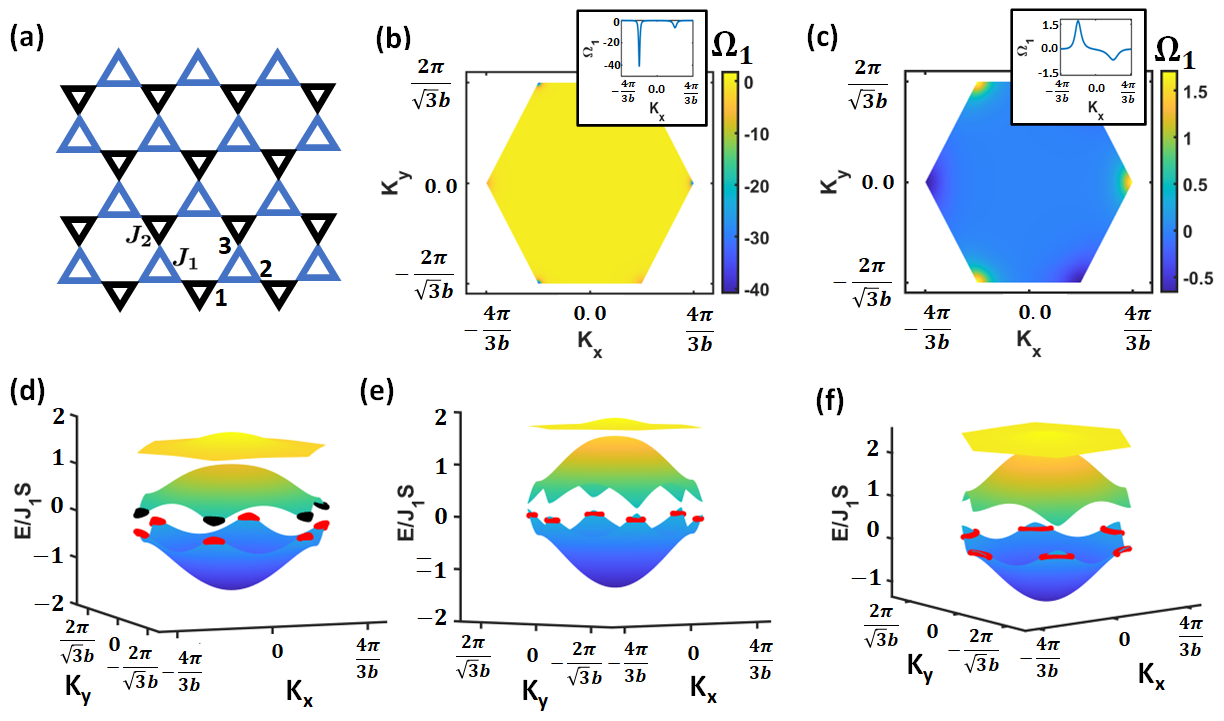} % this command will be ignored
\caption{(a) Schematic of breathing kagome lattice. The Berry-curvature of the lowest magnon band for breathing kagome lattice breathing anisotropies (b) $\delta J=0.05J_1<\delta J^c$, (c) $\delta J=0.5J_1>\delta J^c$. Inset shows the Berry curvature along $K_x$-axis with fixed $K_y=\frac{2\pi}{\sqrt{3b}}$. (d) The band structure for kagome lattice without breathing anisotropy ($\delta J=0$) with external electromagnetic wave with frequency $\Omega=6J_1S$, where the breathing anisotropy $\delta J=0$. The black and red dots represent the points of amplification for the middle and lower band respectively. The band structure of breathing kagome lattice with breathing anisotropies (e) $\delta J=0.05J_1<\delta J^c$, (f) $\delta J=0.5J_1>\delta J^c$ and with external electromagnetic wave with frequency $\Omega=2.7J_1S$. The red dots denote the points of amplification. The other parameters related to electromagnetic coupling strength with the system are $p_0=1.0J_1$, $E_0^x=0.0$, $E_0^y=0.005$, $\gamma=1\times 10^{-4}$.  $\delta J^c=\frac{2}{\sqrt{3}}D$.}
\label{fig::BandAmplification}
\end{figure}
\end{widetext}

\subsection{\label{sec2b} Selective magnon amplification}

Consider the following spin Hamiltonian on a breathing kagome lattice,
\begin{align}
    \pazocal{H}_0=&-J_1\sum_{\left\langle i,j\right\rangle_1} \hat{\boldsymbol{ S}}_i\cdot\hat{\boldsymbol{S}}_j
    -J_2\sum_{\left\langle i,j\right\rangle_2} \hat{\boldsymbol{S}}_i\cdot\hat{\boldsymbol{S}}_j
    \nonumber\\
    &+D\sum_{\left\langle i,j\right\rangle}
    \nu_{ij} \hat{z}\cdot\left(\hat{\boldsymbol{S}}_i\times \hat{\boldsymbol{S}}_j\right)
    -B^z\sum_i \hat{S}_i^z,
    \label{eq::Hamiltonian}
\end{align}
where $\left\langle\dots\right\rangle$ denotes the nearest neighbour (NN) bonds. The subscripts $1$ and $2$ denote the blue and black bonds in Fig.\ref{fig::BandAmplification}(a) respectively, and 
$J_1$ and $J_2$ are the corresponding NN Heisenberg spin-exchange interactions.
$D$ denotes the Dzyaloshinskii-Moriya interaction (DMI) which is chosen to be equal on all NN bonds for simplicity (results are slightly modified by anisotropic DMI, see Appendix\,\ref{appendixC}).
$B^z$ is an external magnetic field perpendicular to the lattice.
We set $J_1=1.0$, $D=0.1J_1$ and $B_z\rightarrow 0^+$ throughout this study;
the parameter $\delta J=J_2-J_1$ denotes the breathing anisotropy.

The ground state of Hamiltonian (\ref{eq::Hamiltonian}) is ferromagnetic for small values of $D$ that are considered here. Low energy magnon excitations above the ground state can be described by the standard Holstein-Primakoff (HP) transformation. In its lowest order, the HP transformation reduces to the linear spin wave transformation defined by $\hat{S}_i^+=\sqrt{2S} \hat{a}_i$, $\hat{S}_i^-=\sqrt{2S}\hat{a}_i^\dagger$, $\hat{S}_i^z=S-\hat{a}_i^\dagger\hat{a}_i$, where $\hat{a}_i^\dagger$ and $\hat{a}_i$ are the magnon creation and annihilation operators respectively. Applying the HP transformation to the Hamiltonian~(\ref{eq::Hamiltonian}) results in the magnon Hmailtonian. Neglecting magnon-magnon interaction terms, and applying Fourier transformation yields a tight-binding magnon Hamiltonian,
\begin{equation}
    \pazocal{H}_0=\sum_{\bk} \Psi_{\bk}^\dagger H_0(\bk) \Psi_{\bk},
    \label{eq::MagnonHamiltonian}
\end{equation}
where $\Psi_{\bk}=\left(\hat{a}_{1,\bk},\,\hat{a}_{2,\bk},\,\hat{a}_{3,\bk}\right)^T$ is the vector of magnon operators in a unit cell and subscripts $1,\,2,\,3$ denote the three basis sites of kagome lattice\,(see Fig.\,\ref{fig::BandAmplification}(a)).
$\pazocal{H}_0(\bk)$ is the non-interacting magnon Hamiltonian  matrix ( Appendix\,\ref{appendixB}).

The (non-interacting) magnon bands are obtained by diagonalizing $\pazocal{H}_0(\bk)$. For the regular kagome lattice without the breathing anisotropy\,($\delta J=0$) and any DMI\,($D=0$), the magnon spectrum consists of three bands - dispersive lower and middle bands and a dispersionless (flat) upper band. The lower and middle bands touch at Dirac points at the corners of the Brillouin zone; the middle and upper bands touch at a quadratic band touching point at the zone center. For finite DMI\,($D\neq 0$) or breathing anisotropy\,($\delta J\neq 0$), a band gap opens between the lower and middle bands at $K$ and $K'$ points \,(Figs.\,\ref{fig::BandAmplification}(d)-(f)).
%and the bands acquire  topological nature.
%Depending on the type of factor\,(either breathing anisotropy or DMI) that opens the band, the Berry-curvature distribution of bands as well as Chern number will differ, where the expression of Chern number $C_n$ and Berry curvature $\Omega_n(\bk)$ for $n$-th band are given by,
%
%three magnon bands are obtained\,(see Fig.\,\ref{fig::BandAmplification}(d), (e) and %(f)).
%{
%\begin{equation}
%    \Omega^z_n(\bk) \!=\! i\sum_{m\neq n} 
%    \frac{\squeezeD{m(\bk)}{\frac{\partial \pazocal{H}}{\partial k_x}}{n(\bk)}\! 
%    \squeezeD{m(\bk)}{\frac{\partial \pazocal{H}}{\partial k_y}}{n(\bk)}-(k_x \!\leftrightarrow \! k_y)}{(E_n(\bk)-E_m(\bk))^2},
%    \label{BerryCurvature}
%\end{equation}
%}
%{\begin{equation}
%\resizebox{1\linewidth}{!}{$ \Omega^z_n(\bk) = {\small i\sum_{m\neq n}} \frac{\left\langle m(\bk)\right|\frac{\partial \pazocal{H}}{\partial k_x}\left|n(\bk)\right\rangle \left\langle n(\bk)\right|\frac{\partial \pazocal{H}}{\partial k_x}\left|m(\bk)\right\rangle-(k_x \leftrightarrow  k_y)}{\left[E_n(\bk)-E_m(\bk)\right]^2}, $}
    \label{BerryCurvature}
%\end{equation}
%}
%\begin{equation}
%    C_n=\frac{1}{2\pi}\int_{\text{BZ}} d\bk \Omega_n(\bk),
%\end{equation}
%\begin{equation}
%    \resizebox{1\linewidth}{!}{$\Omega_n(\bk) = i\sum\limits_{m\neq n} 
%    \frac{\squeezeD{m(\bk)}{\frac{\partial \pazocal{H}_0}{\partial k_x}}{n(\bk)} 
%    \squeezeD{m(\bk)}{\frac{\partial \pazocal{H}_0}{\partial k_y}}{n(\bk)}-(k_x \leftrightarrow k_y)}{(E_n(\bk)-E_m(\bk))^2},$}
%    \label{BerryCurvature}
%\end{equation}
%where $\ket{m(\bk)}$ and $E_m(\bk)$ are the eigenstate and eigenvalue of $m$-th band at $\bk$-point.
The Berry curvature distribution 
%and the topology 
of the lower and middle magnon bands as a function of breathing anisotropy and DMI can be obtained using the linearized effective Hamiltonian near $K$ and $K'$ points\,\cite{KP_Theory_1, KP_Theory_2, KP_Theory_3}%\,\cite{KdotPTheory_Note},
\begin{equation}
    \pazocal{H}_{\text{eff}}=\alpha v \left( k_x^\prime \sigma_z + k_y^\prime \sigma_x \right)+ m_\alpha \sigma_y,
\end{equation}
where, $v=\frac{\sqrt{3}}{2} J_1 S$ and $m_\alpha=S\left[ \frac{1}{\sqrt{3}} D + \alpha \frac{3}{2}\delta J\right]$;  $\alpha=+1$ ($-1$) at the $K$ ($K'$) point.
The mass term $m_\alpha$, which is instrumental in opening up the band gap, depends on both the DMI, $D$, and the breathing anisotropy, $\delta J$.
Interestingly, the signs of $\delta J$ in $m_\alpha$ at $K$ and $K'$ are opposite, and hence the breathing anisotropy would result in opposite Berry curvatures at $K$ and $K'$ points of the same band.
%, leading to a net zero Chern number.
%Whereas, the effect of DMI at $K$ and $K'$ point is equal, resulting in equal Berry curvature at these points, leading to non-zero Chrern number or topologically non-trivial band.
In contrast, DMI has an equal effect in the mass term at $K$ and $K'$ points, resulting in an equal Berry curvature at these two points of the same band.
%and a non-zero Chern number or topologically non-trivial band.
In the presence of both DMI and breathing anisotropy, there is a band topological phase transition when the relative strength of the two is varied.  
At zero or small values of $\delta J$, the DMI contribution is dominant and the lowest band features the same sign of Berry curvature around $K$ and $K'$ points\,(see Fig.\,\ref{fig::BandAmplification}(b)).
%Chern numbers. 
For $\delta J$ larger than a critical value $\delta J^c=\frac{2}{\sqrt{3}}D$, the mass term acquires opposite signs at $K$ and $K'$ points, and the total Berry curvature vanishes. Locally, the Berry curvature is opposite in sign at the $K$ and $K'$ points, with inequivalent distribution\,(see Fig.\,\ref{fig::BandAmplification}(c)).
The magnitude of Berry curvature diminishes as $\delta J$ increases for $\delta J> \delta J^c$.
%and the net Chern number vanishes, yielding topologically trivial bands (but with non-trivial Berry curvature distribution). 
As discussed below, this plays a crucial role in amplification of bulk magnon modes in isolated bands.
The Berry curvatures for $\delta J<\delta J^c$ and $\delta J>\delta J^c$ are plotted in Figs.\,\ref{fig::BandAmplification}(b) and \ref{fig::BandAmplification}(c) respectively.
%In this study, we considered $D>D_c$, so that the Berry curvature at $K$ and $K'$ points are equal\,(see Fig.\,\ref{fig::BandAmplification}(b)) and thus selectively amplifying magnons nearby $K$ and $K'$ points would result in a amplification of thermal Hall conductance. 
The mechanism of amplification due to coupling to an external EM field is discussed next.

Electromagnetic field couples to a magnetic insulator via the polarization operator\,\cite{PolarizationOperator1,PolarizationOperator2,PolarizationOperator3}. 
Both onsite and NN-bond polarization operators are allowed by the symmetry of breathing kagome magnet.
The onsite polarization is associated with linear Stark effect\,\cite{Moriya} whereas the NN-bond polarization introduces a virtual electron hopping\,\cite{MagnonAmplification}.
To minimize the number of parameters we restricted the polarization operator on NN bonds as considered in Ref.\,\cite{MagnonAmplification}. Thus, the minimal microscopic Hamiltonian describing the coupling between an EM field and a magnetic insulator is given by,
\begin{align}
     \pazocal{H}_c &=\cos(\Omega t)\boldsymbol{E}\cdot\sum_{\left\langle i,j\right\rangle} \boldsymbol{P}_{ij}
     \nonumber\\
     \text{with}, \mathbf{P}_{ij}&\approx \boldsymbol{p}_{0,ij} \left(\mathbf{S}_i\cdot\mathbf{Q}_{ij}\right)
    \left(\mathbf{S}_j\cdot\mathbf{Q}_{ij}\right).
    \label{eq::Polarization}
 \end{align}
 $\boldsymbol{E}$ is the electric field amplitude of an external electromagnetic field; $\mathbf{p}_{0,ij}$ and $\mathbf{Q}_{ij}$ are defined in Appendix\,\ref{appendixA}.
 Other polarization terms are neglected as they will be eliminated in the rotating wave approximation due to absence of magnon pair creation and annihilation operators.
 
 The total Hamiltonian for a system with toroidal boundary condition for magnons coupled to an external EM field takes the form,
\begin{align}
    \pazocal{H}=&\pazocal{H}_0+\pazocal{H}_c
    \nonumber\\
    =&\frac{1}{2}\sum_{\mathbf{k}}
    \begin{pmatrix}
    \Psi_{\bk}^\dagger & \Psi_{-\bk}
    \end{pmatrix}
    \begin{pmatrix}
    H_0(\bk) & O_{3\times 3} \\
    O_{3\times 3} & H_0(-\bk)^T
    \end{pmatrix}
    \begin{pmatrix}
    \Psi_{\bk} \\ \Psi_{-\bk}^\dagger
    \end{pmatrix}
    \nonumber\\
    &+
    \frac{1}{2}\cos(\Omega t)\sum_{k} 
    \begin{pmatrix}
    \Psi_{\bk}^\dagger & \Psi_{-\bk}
    \end{pmatrix}
    \begin{pmatrix}
    O_{3\times 3} & H_c(\bk) \\
    H_c(\bk)^\dagger & O_{3\times 3}
    \end{pmatrix}
    \begin{pmatrix}
    \Psi_{\bk} \\ \Psi_{-\bk}^\dagger
    \end{pmatrix},
    \label{eq::TotalHamiltonian}
\end{align}
where $O_{3\times 3}$ is a null matrix of size $3\times 3$.
%The first matrix is derived from the unperturbed Hamiltonian Eq.\,\ref{eq::Magnon} and the second matrix is derived from the coupling Hamiltonian Eq.\,\ref{eq::Coupling}.
The expressions for the matrices $H_0(\bk)$ and $H_c(\bk)$ are given in Appendix\,\ref{appendixB}. Let $\tilde{\Psi}_\bk (\tilde{\Psi}_{-\bk}^\dagger)$ denote the eigenvectors of $H_0(\bk)\,(H_0(-\bk)^T)$ and $U_1(\bk)(U_2(\bk))$ be the corresponding unitary transformation relating them to the original basis: $\tilde{\Psi}_\bk=U_1(\bk)\Psi_\bk$, and $\tilde{\Psi}_{-\bk}^\dagger=U_2(\bk)\Psi_{-\bk}^\dagger$. The Hamiltonian $\pazocal{H}$ is first expressed in the basis of $\tilde{\Psi}_\bk$ and $\tilde{\Psi}_{-\bk}^\dagger$ and then transformed from the lab-frame to rotating-frame by the unitary operator $U(t)=\exp{\frac{i\omega t}{2}\sum_\bk  \tilde{\Psi}_\bk^\dagger \tilde{\Psi}_\bk}$. Neglecting the time-dependent terms, we get the following effective Hamiltonian in the rotating frame,
%First representing the Hamiltonian $\pazocal{H}$ in the diagonal basis $\tilde{\Psi}_\bk=U_1(\bk)\Psi_\bk$, $\tilde{\Psi}_{-\bk}^\dagger=U_2(\bk)\Psi_{-\bk}^\dagger$ of matrices $H_0(\bk)$, $H_0(-\bk)^T$; and then transforming the system from the lab-frame to rotating-frame by using the unitary operator $U(t)=\exp{\frac{i\omega t}{2}\sum_\bk  \tilde{\Psi}_\bk^\dagger \tilde{\Psi}_\bk}$; and neglecting the time-dependent terms, we get the following effective Hamiltonian,
\begin{equation}
\pazocal{H}_{\text{eff}}=\frac{1}{2}
\sum_{k}
\begin{pmatrix}
\tilde{\Psi}_\bk^\dagger  & \tilde{\Psi}_{-\bk}
\end{pmatrix}
\begin{pmatrix}
\epsilon_\bk-\frac{\Omega}{2} & \frac{\tilde{H}_{c}(\bk)}{2} \\
\frac{\tilde{H}_{c}(\bk)}{2} & \epsilon_{-\bk}-\frac{\Omega}{2}
\end{pmatrix}
\begin{pmatrix}
\tilde{\Psi}_\bk  \\ \tilde{\Psi}_{-\bk}^\dagger
\end{pmatrix},
\label{eq::EffectiveHamiltonian}
\end{equation}
where $\epsilon_{\bk}$ and $\epsilon_{-\bk}$ are the diagonal matrices of eigenvalues of $H_0(\bk)$ and $H_0(-\bk)^T$ respectively.
$\tilde{H}_c(\bk)=U_1 H_c(\bk) U_2^\dagger$ is the coupling Hamiltonian in the diagonal basis. 
We note that even if the coupling matrix in Eq.\,\ref{eq::TotalHamiltonian} contained any diagonal terms representing magnon hopping, they would not appear in the effective Hamiltonian due to the rotating wave approximation. Consequently, only the terms in the polarization operator that result in pair-creation and annihilation are considered in Eq.\,\ref{eq::Polarization}.
The equation of motion of the field $\left\langle \tilde{\Psi}_{\bk}\right\rangle=\tilde{\alpha}_\bk=\left(\left\langle\hat{\tilde{a}}_{1,\bk}\right\rangle,\,\left\langle\hat{\tilde{a}}_{2,\bk}\right\rangle,\,\left\langle\hat{\tilde{a}}_{3,\bk}\right\rangle\right)^T$ is given by,
{\small
\begin{equation}
    \frac{d}{dt}
    \begin{pmatrix}
    \tilde{\alpha}_\bk^*\\ 
    \tilde{\alpha}_{-\bk}
    \end{pmatrix}
    =
    \mathrm{i}
    \begin{pmatrix}
    \tilde{\epsilon}_\bk-\mathrm{i}\frac{\gamma\mathbb{I}+\eta\left|\alpha_k\right|^2}{2} & 
    \frac{\tilde{H}_c(\bk)}{2} \\
    -\frac{\tilde{H}_c(\bk)}{2} &
    -\tilde{\epsilon}_{-\bk}-\mathrm{i}\frac{\gamma\mathbb{I}+\eta\left|\alpha_\bk\right|^2}{2}
    \end{pmatrix}
    \begin{pmatrix}
    \tilde{\alpha}_\bk^*\\ 
    \tilde{\alpha}_{-\bk}
    \end{pmatrix},
    \label{eq::EOM}
\end{equation}
}where $\tilde{\epsilon}_\bk$ and $\tilde{\epsilon}_{-\bk}$ are the diagonal matrix of eigenvalues of matrix $H_0(\bk)$ and $H_0(-\bk)^T$ with $\frac{\Omega}{2}$ subtracted from them; 
$\gamma$ and $\eta$ are phenomenological constants included to describe linear and non-linear damping respectively; 
$\mathbb{I}$ is $3\times 3$ identity matrix and $\left|\alpha_\bk\right|^2$ is diagonal matrix with entries {\small $\left(\left|\left\langle\hat{\tilde{a}}_{1,\bk}\right\rangle\right|^2,\,\left|\left\langle\hat{\tilde{a}}_{2,\bk}\right\rangle\right|^2,\,\left|\left\langle\hat{\tilde{a}}_{3,\bk}\right\rangle\right|^2\right)$}.

The dynamical matrix governing the time evolution of $\left\langle \tilde{\Psi}_{\bk}\right\rangle$ feature complex eigenvalues (for $\eta=0$) -- 
%square matrix at the right hand side of Eq.\,\ref{eq::EOM} is dynamical matrix and the eigenvalues of the matrix (considering $\eta=0$) is complex.
the real and imaginary parts represent the energy and lifetime of a magnon.
In the absence of electromagnetic coupling (or {\footnotesize $\tilde{H}_c(\bk)\approx O_{3\times 3}$}), the imaginary part of eigenvalues is negative indicating magnon decay.
However, as the amplitude of the EM field increases, the imaginary part of some of the eigenvalues satisfying $\epsilon_\bk+\epsilon_{-\bk}= \Omega$ become positive denoting spontaneous amplification of the number of magnons.

\begin{figure}[htb]
\includegraphics[width=0.5\textwidth]{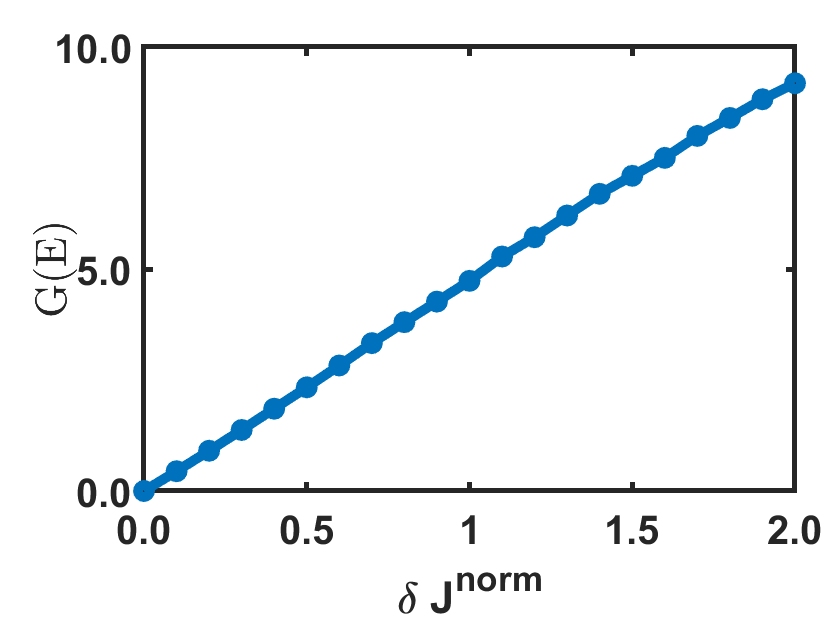} % this command will be ignored
\caption{ The strength of coupling between the magnons and EM field $G(E)$ as a function of normalized breathing anisotropy $\delta J^{\text{norm}}=\delta J/\delta J^c$ for the lowest band near energy $E=2.7J_1S/2$ (frequency of light $\Omega=2.7J_1S$). The parameters for the coupling $p_0=1.0J_1$, $E_0^x=0.0$, $E_0^y=0.005$.
\label{fig::CouplingConstantVsDeltaJ}
}
\end{figure}

The coupling between the system and electromagnetic wave depends not just on the amplitude of the wave, but also the coupling constant $\boldsymbol{p}_0$ and breathing anisotropy $\delta J$. Note that, the effect of breathing anisotropy is only considered in the Heisenberg interaction term Eq.\,\ref{eq::Hamiltonian}; it does not appear in the coupling term Eq.\,\ref{eq::Polarization}, but it affects the 
 coupling terms $\tilde{H}_c(\bk)$ in Eq.\ref{eq::EOM} through the unitary matrices 
 $U_1(\bk)$ and $U_2(\bk)$ that define the change of basis of $\pazocal{H}$ from 
 $\begin{pmatrix}\Psi_{\bk}^\dagger & \Psi_{-\bk}\end{pmatrix}^T$  to the eigenbasis of $H_0(\bk)$ and $H_0(-\bk)^T$, viz., 
 $\begin{pmatrix}\tilde \Psi_{\bk}^\dagger & \tilde\Psi_{-\bk}\end{pmatrix}^T$.
The  strength of coupling, $G(E)$, between the EM field and pairs of magnons (defined as the sum of matrix elements $[\tilde{H}_c(\bk)]_{11}$ over Brillouin zone) in the lowest band with energy $E\approx 2.7J_1S/2$  is shown as a function of $\delta J$ in Fig.\,\ref{fig::CouplingConstantVsDeltaJ}. The figure illustrates why the original scheme of Malz, et.al. cannot amplify bulk magnon modes in an isolated band in the presence of inversion symmetry. The inversion symmetry of the regular kagome lattice prevents selective amplification of magnons with the same energy.
Instead, magnons in pair of bands are simultaneously  amplified by an external EM field with frequency close to the middle of the gap between two magnon bands as shown in Fig.\,\ref{fig::BandAmplification}(d),
% In Fig.\,\ref{fig::BandAmplification}(d), 
where amplified magnons from middle and lower bands
%with positive imaginary parts of eigenvalues 
are marked by black and red dots, respectively. 

 The breathing anisotropy breaks inversion symmetry and, in turn, allows for the generation of field induced magnons at any energy within a single band. Our results indicate that the amplification increases linearly with the breathing anisotropy.  The energy of magnon amplification points can be adjusted by tuning the frequency of the external electromagnetic field.
 In Figs.\,\ref{fig::BandAmplification}(e) and \ref{fig::BandAmplification}(f), the magnons of the lowest band of breathing kagome lattice are amplified near $K$ and $K'$ points (at energies $E=2.7J_1S/2$). 
 Due to finite Berry-curvatures near $K$ and $K'$ points\,(see Figs.\,\ref{fig::BandAmplification}(b), \ref{fig::BandAmplification}(c)), we expect an amplification of thermal Hall conductivity, which is discussed in the next section.

 \begin{widetext}
 
 \begin{figure}[htb]
\includegraphics[width=\textwidth]{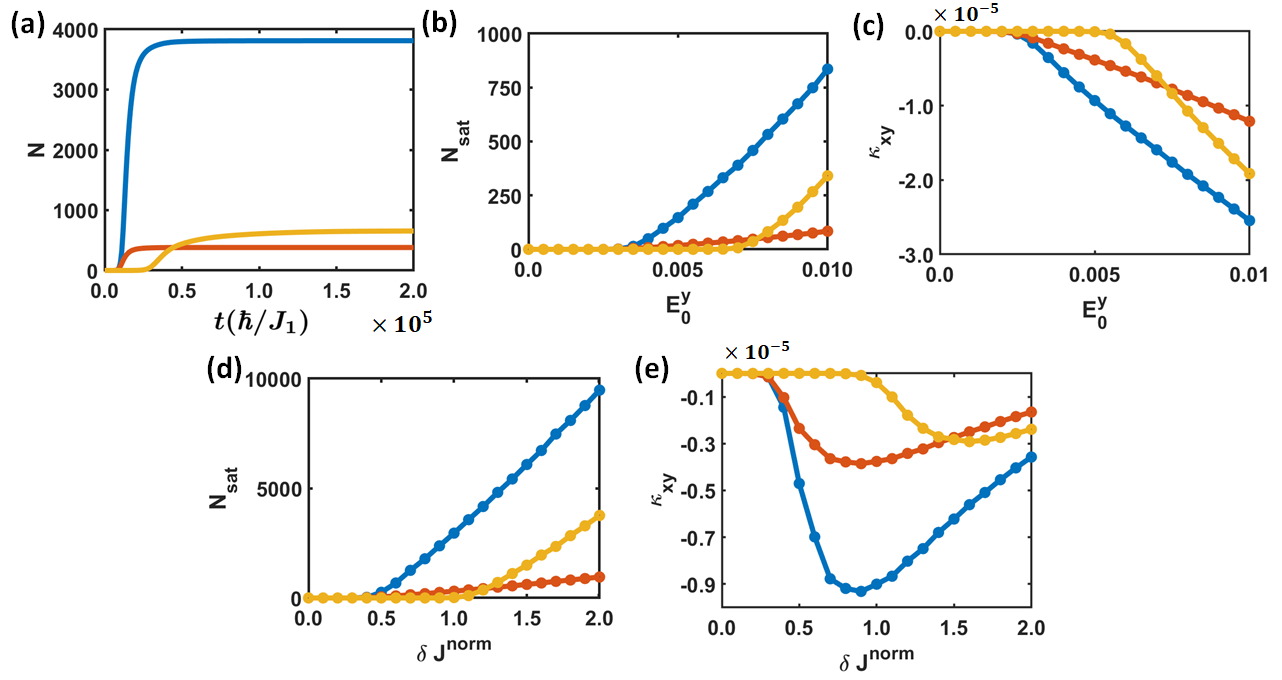} % this command will be ignored
\caption{(a) Total numbers of amplified magnons as a function of time. (b) The total number of magnons at saturation as a function of electric field amplitude $E_0^y$. (c) The thermal Hall conductance as a function of electric field amplitude $E_0^y$. (d) The total number of magnons at saturation as a function of normalized breathing anisotropy $\delta J^{\text{norm}}=\delta J/\delta J^c$ . (e)  The thermal Hall conductance as a function of normalized breathing anisotropy $\delta J^{\text{norm}}=\delta J/\delta J^c$ . The parameters of the system and electromagnetic field, if it is not varied, are $\delta J=0.1J_1$, $p_0=1.0J_1$, $E_0^x=0.0$, $E_0^y=0.005$, $T=0.1J_1$, $\Omega=2.7J_1S$. The values of linear and non-linear dampings for blue, red and yellow curves are $(\gamma=10^{-4}, \eta=10^{-3})$, $(\gamma=10^{-4}, \eta=10^{-2})$, $(\gamma=10^{-3}, \eta=10^{-3})$ respectively. The system size for the simulations is $3\times 5000 \times 5000$, where $5000$ denotes the number of unit cells along each translation vector.}
\label{fig::HallAmplification}
\end{figure}

\end{widetext}

\subsection{\label{sec2c} Amplification of thermal Hall conductivity}
Solution of the equation Eq.\,\ref{eq::EOM} gives the magnon population as a function of time at different $\bk$-points.
Fig.\,\ref{fig::HallAmplification}(a) shows the growth of the total number of magnons N$=\sum_{\bk}\left|\alpha_{1,\bk}\right|^2$ with time for various damping parameters.
%As a result of nonlinear damping $\eta$\,\cite{MagnonAmplification}, 
The number of amplified magnons saturates to a finite value N$_\text{sat}$ and the system reaches a steady state due to the nonlinear damping denoted by the parameter $\eta$\,\cite{MagnonAmplification}. The {\it number} of magnons at saturation and the time to reach saturation depends on both $\gamma$ and $\eta$.
 The non-linear damping arises due to 
%The number of magnons saturates due to nonlinear damping which is a by product of 
magnon-magnon, magnon-phonon and magnon-impurity interactions, whereas the linear damping parameter, $\gamma$, captures the decoherence of amplified magnons. Strictly speaking, the various interaction terms also affect the magnon band structure, but any such band structure renormalization is not considered in this study.
%the effect of interaction on band structure is not considered in this study.
Consequently, the amplification mechanism near saturation is not captured in this study, and the observable thermal Hall conductance is calculated at half of the saturation value N$_{\text{sat}}$ and treated as a transient observable.

Fig.\,\ref{fig::HallAmplification}(b) illustrates the total magnon population at saturation N$_{\text{sat}}$ as a function of the electric field amplitude $E_0^y$ for different damping parameters. At small values of $E_0^y$, there are no amplified magnons. The amplitude of electric field (equivalently the intensity of the incident EM field) has to exceed a threshold before the onset of magnon amplification. Once that is crossed, 
there is a linear increase in amplified magnons with $E_0^y$.
 The threshold field amplitude depends on the linear damping $\gamma$\,\cite{MagnonAmplification}, whereas the rate of increase in amplification with respect to amplitude (slope of the plots in Fig.\,\ref{fig::HallAmplification}(b)) depends on the non-linear damping $\eta$.

Next we calculate the thermal Hall conductance as a function of amplitude of light as in Fig.\ref{fig::HallAmplification}(c) and the expression of dimensionless thermal Hall conductance is given by\,\cite{HallEffect1,HallEffect2,HallEffect3,HallEffect4},
\begin{equation}
    \kappa_{xy}=\frac{T}{A} \sum_\bk \sum_{n=1}^N \left\lbrace c_2\left[\rho_{n,\bk}\right]-\frac{\pi^2}{3}\right\rbrace
\Omega_n(\bk),
\label{eq::ThermalHallConductance}
\end{equation}
where A and T are the area and dimensionless temperature of the system respectively; $\rho_{n,\bk}$ is the number of magnons with momentum $\bk$ in the $n$-th band.
thermal excitations as well as amplification both contribute to $\rho_{n,\bk}$. 
The function $c_2(x)$ is given by $c_2(x)=(1+x)(\ln\frac{1+x}{x})^2-(\ln x)^2-2\text{Li}_2(-x)$ with $\text{Li}_2(x)$ as bilogarithmic function.
The thermal Hall conductance grows linearly as the amplitude of light increases due to linear increase in the number of magnons.
%This feature is due to the nature of functional $\lim\limits_{x\to 0}c_2(x)\propto x$ in the expression of thermal Hall conductance Eq.\,\ref{eq::ThermalHallConductance}.

The dependence of the total number of magnons at saturation N$_{\text{sat}}$ on breathing anisotropy is shown in Fig.\,\ref{fig::HallAmplification}(d).
The figure shows that the number magnons at saturation increases linearly as the breathing anisotropy $\delta J$ increases, which is expected as the coupling between the light and the system also increases as a function of $\delta J$ as in Fig.\,\ref{fig::CouplingConstantVsDeltaJ}.
The qualitative nature of the plots and the dependence on the damping parameters are same as in Fig.\,\ref{fig::HallAmplification}(c), because both of the amplitude of light and the breathing anisotropy increase the coupling amplitude between the system and electromagnetic field.

Finally, Fig.\,\ref{fig::HallAmplification}(e) illustrates the thermal Hall conductance $\kappa_{\text{xy}}$ as a function of breathing anisotropy,  showing that the thermal Hall conductance varies non-monotonically with $\delta J$ -- 
%, unlike the $E_0^y$ dependence\,(see Fig.\,\ref{fig::HallAmplification}(c));  
it remains zero up to some threshold $\delta J$ (that depends on the linear damping, $\gamma$), and then increases rapidly in magnitude with increasing breathing anisotropy, reaching a maximum at an intermediate values of $\delta J$. The position and height of the conductance maximum depends on both linear and non-linear damping quantified by $\gamma$ and $\eta$.  With further increase in $\delta J$, the thermal Hall conductance decreases monotonically. 
This can be understood as follows. Varying $\delta J$, not only increases coupling between system and electromagnetic wave as in Fig.\,\ref{fig::CouplingConstantVsDeltaJ}, but also changes the Berry curvature distribution. In the limit $\delta J\rightarrow 0$ the Berry-curvature at $K$ and $K'$ points are equal in magnitude and sign, whereas in the limit $\delta J\gg D$ the Berry curvature at $K$ and $K'$ points vanish\,(See Sec.\,\ref{sec2b} and Fig.\,\ref{fig::BandAmplification}). 
Thus in both the limits $\delta J\rightarrow 0$ and $\delta J\gg D$, the enhancement of thermal Hall conductance is expected to vanish; the thermal Hall conductance reaches a maximum at finite $\delta J\sim \delta J^c$ as in Fig.\,\ref{fig::HallAmplification}(e).

The thermal Hall conductivity without amplification is $\kappa_{\text{xy}}=-1.7\times 10^{-8}$ for the system parameters considered in Fig.\,\ref{fig::HallAmplification}; the results in Figs.\,\ref{fig::HallAmplification}(c) and \ref{fig::HallAmplification}(d) show that the magnon thermal Hall conductivity can be amplified by three orders of magnitude\,\cite{MagnonHallEffect,MagnonHallEffect2,MagnonHallEffect3}.
Thus selective amplification of magnons is a promising way for amplification and detection of magnon thermal Hall effect.

 The thermal Hall conductance that is calculated and plotted in this paper is for a pure two-dimensional system. Moreover, we use dimensionless temperature and thermal Hall conductance for convenience. For interested experimental readers, we reproduce some data points in conventional units, which also clarifies whether thermal Hall conductance can be measured within experimental accuracy. The relation between experimentally measured thermal Hall conductance and temperature with their dimensionless counterpart are respectively\,\cite{WeylTriplon},
\begin{equation}
    \kappa^{\text{3D}}_{xy}=\kappa_{xy} \frac{N_z k_B J_1}{l_z\hbar},\,
    T^\prime=\frac{T}{k_B},
\end{equation}
where, $N_z$ and $l_z$ are the number of 2D layers and the dimension of sample in z-direction respectively. 
Moreover, $k_B$ and $\hbar$ are Boltzmann constant and Planck's constant respectively.
We consider the following ideal values of system parameters to calculate the observable, $l_z=1cm$, $N_z=10^8$, $J_1=1meV$.                                  Throughout this study we use the temperature $T=0.1J_1$ which corresponds to $T'\approx 1K$.
Moreover, the order of magnitude of Hall conductivity with and without amplification are $\kappa_{xy}=10^{-5}$ and $\kappa_{xy}=10^{-8}$ which correspond to the Hall conductivity $\kappa_{xy}^{3D}\approx 10^{-4} W/Km$ and $\kappa_{xy}^{3D}\approx 10^{-7} W/Km$ respectively.
Consequently, the thermal Hall conductance can be measured within experimental accuracy ($10^{-4}$ to $10^{-3}$ W/Km) only with amplification in this example system.\,\cite{noTHE, MagnonHallEffect}.

\section{Conclusion}
In conclusion, we have demonstrated that it is possible to use an external EM field to excite magnons at any chosen energy in a single band  by breaking inversion symmetry of the magnetic lattice.
In particular, the selective amplification scheme can be used to amplify the magnons at finite Berry curvature and amplify thermal Hall signal in quantum magnets with broken inversion symmetry and topological magnon bands. Although we are not aware of any real quantum magnet satisfying all the necessary conditions, there are several promising possibilities. The metallic chiral magnet \ce{Gd2Ru4Al12} has a breathing kagome magnetic lattice with ferromagnetic near neighbor interactions, but negligible native DMI. Inducing DMI externally can impart finite Berry curvature to the magnon bands that can then be selectively amplified according to the protocol presented here. Conversely, if inversion symmetry can be externally broken in materials such as the Kagome ferromagnets Haydeeite\,\cite{KagomeMaterial2} and Cu(1-3, bdc)\,\cite{Magnon4,TopologicalMagnons7}, \ce{NaV3(OH)6(SO4)2}\,\cite{KagomeMaterial3}, the honeycomb ferromagnets, \ce{CrI3}\,\cite{CrI3_1,CrI3_2,CrI3_3}, \ce{CrBr3}\,\cite{CrBr3_1,CrBr3_2}, \ce{CrCl3}\,\cite{CrCl3_1,CrCl3_2,CrCl3_3}, \ce{CrGeTe3}\,\cite{CrGeTe3}, \ce{CrSiTe3}\,\cite{CrSiTe3_1}, or the Shastry-Sutherland compound, \ce{SrCu2(BO3)2}\,\cite{SrCu2(BO3)2_V1,SrCu2(BO3)2_V2,WeylTriplon} that are known to host topological magnon bands, the present scheme can be used to induce finite thermal Hall effect that has so far eluded experiments\,\cite{noTHE,noTHE2}.
However, in this study, we have not included higher order topological edge states or hinge states, which might be of interest in the future\,\cite{HingeState,HingeState2,HingeState3}.
Overall, we believe our results will stimulate further theoretical and experimental investigations in this direction.

%subsequently the thermal Hall signal can be amplified two order of magnitude higher than the usual value.
%Our work has illustrated that the magnons can be selectively amplified for the breathing kagome ferromagnet. Although at present no material is known to be a breathing Kagome ferromagnet, the selective topological magnon amplification scheme can still be applied to the Kagome ferromagnets Haydeeite\,\cite{KagomeMaterial2} and Cu(1-3, bdc)\,\cite{KagomeMaterial1_V1,KagomeMaterial1_V2}, \ce{NaV3(OH)6(SO4)2}\,\cite{KagomeMaterial3} by introducing breathing anisotropy due to application of stress.
%We hope that our findings will also stimulate theoretical and experimental investigations of the topological signatures of magnons in honeycomb ferromagnets \ce{CrI3}\,\cite{CrI3_1,CrI3_2,CrI3_3}, \ce{CrBr3}\,\cite{CrBr3_1,CrBr3_2,CrBr3_3}, \ce{CrCl3}\,\cite{CrCl3_1,CrCl3_2,CrCl3_3}, \ce{CrGeTe3}\,\cite{CrGeTe3}, \ce{CrSiTe3}\,\cite{CrSiTe3_1,CrSiTe3_2} which are bosonic analogue of Haldane model\,\cite{Haldane}.
%Furthermore the dimerized Shastry-Sutherland compound \ce{SrCu2(BO3)2}\,\cite{SrCu2(BO3)2_V1,SrCu2(BO3)2_V2,WeylTriplon} has long been proposed to be host of topological triplons, but experimental studies has failed to measure any thermal Hall response\,\cite{noTHE,noTHE2}.
%Thus our scheme to selectively amplify magnons with finite Berry-curvature in reciprocal space is a promising tool to be successful in measuring a finite thermal Hall effect in a magnetic systems.

\begin{acknowledgements}
Y.B would like to acknowledge the support by the National Research Foundation, Singapore under the NRF fellowship award (NRF-NRFF12-2020-005). PS acknowledges financial support from the Ministry of Education, Singapore through MOE2019-T2-002-119.
\end{acknowledgements}

\appendix
\begin{widetext}

\section{\label{appendixA} Polarization Operator}

In this appendix, the mathematical details of the polarization term are discussed. The polarization operator is given as,
\begin{equation}
     \mathbf{P}_{ij}\approx \boldsymbol{p}_{0,ij} \left(\mathbf{S}_i\cdot\mathbf{Q}_{ij}\right)
    \left(\mathbf{S}_j\cdot\mathbf{Q}_{ij}\right),
 \end{equation}
 which is derived from the following Hubbard model of electrons on kagome lattice,
 \begin{equation}
      \pazocal{H}_{\text{Hubbard}}
      =
      -\sum_{ij}
      \left[
      \begin{pmatrix}
      \hat{c}_{i\uparrow}^\dagger &
      \hat{c}_{i\downarrow}^\dagger
      \end{pmatrix}
      \left(t\mathbb{I}\cos(\theta)+it\boldsymbol{n}\cdot\boldsymbol{\sigma}\sin(\theta)\right)
      \begin{pmatrix}
      \hat{c}_{j\uparrow} \\
      \hat{c}_{j\downarrow}
      \end{pmatrix}
      +
      \text{H.c.}
      \right]
      +
      U\sum_i \hat{n}_{i\uparrow}\hat{n}_{j\downarrow}
 \end{equation}
 where $t\cos(\theta)$ and $it\boldsymbol{n}\sin(\theta)$ ($\boldsymbol{n}$ is a unit vector) are the real and complex hopping amplitudes of electrons on nearest neighbour bonds respectively. $U$ is the onsite Coulomb repulsion. The $\bold{p}_{0,ij}$ and $\boldsymbol{Q}_{ij}$ are given by,
\begin{align*}
    \boldsymbol{p}_{0,ij} &=-16\theta^2 e a \frac{t^3}{U^3} (\boldsymbol{e}_{jk}-\boldsymbol{e}_{ki})
    =p_0 (\boldsymbol{e}_{jk}-\boldsymbol{e}_{ki})
    ,
    \\
    \boldsymbol{Q}_{ij}&=\boldsymbol{n}-n^z\hat{z}
\end{align*}
where $e$ and $a$ are the electron charge and lattice constant respectively.
$\boldsymbol{e}_{jk}$ is a vector on nearest neighbour bonds from site-$j$ to site-$k$.
The sites $i$, $j$ and $k$ are the sites on the same triangle of the kagome lattice.

\section{\label{appendixB} Matrices $H_0(\bk)$ and $H_c(\bk)$}
The matrices $H_0(\bk)$ and $H_c(\bk)$ which appeared in equations  Eq.\,\ref{eq::TotalHamiltonian} and Eq.\,\ref{eq::MagnonHamiltonian} are explicitly given here,

\begin{equation}
    H_0(\bk)=
\begin{pmatrix}
    \epsilon & F(\boldsymbol{a})^* & F(\boldsymbol{b}) \\
    F(\boldsymbol{a}) & \epsilon & F(\boldsymbol{c}) \\
    F(\boldsymbol{b})^* & F(\boldsymbol{c})^* & \epsilon
\end{pmatrix},
\quad
H_c(\bk)=\begin{pmatrix}
    0 & F_{12}(-\bk) & F_{13}(\bk) \\
    F_{12}(\bk) & 0 & -F_{23}(-\bk) \\
    F_{13}(-\bk) & -F_{23}(\bk) & 0,
\end{pmatrix}
\end{equation}
where, $\epsilon=2\left(J_1+J_2\right)S+B_z$; $F(\boldsymbol{x})=-S\left(J_1 e^{i\bk\cdot\frac{\boldsymbol{x}}{2}}+J_2 e^{-i\bk\cdot\frac{\boldsymbol{x}}{2}}\right)\pm S\left(J_1 e^{i\bk\cdot\frac{\boldsymbol{x}}{2}}+J_2 e^{-i\bk\cdot\frac{\boldsymbol{x}}{2}}\right)$ and the positive sign is for $\boldsymbol{x}=\boldsymbol{a},\boldsymbol{b}$ and the negative sign is for $\boldsymbol{x}=\boldsymbol{c}$, where the lattice transnational vectors $\boldsymbol{a}=(1,0)$, $\boldsymbol{b}=\left(-\frac{1}{2},-\frac{\sqrt{3}}{2}\right)$ and $\boldsymbol{c}=\boldsymbol{a}+\boldsymbol{b}$
; $F_{pq}(\bk)=\left[c_{pq} e^{i\bk\cdot\frac{\boldsymbol{a}_{pq}}{2}}-c^\prime_{pq} e^{-i\bk\cdot\frac{\boldsymbol{a}_{pq}}{2}}\right]$, where $c_{ij}=c_{ji}=-p_0\left(\boldsymbol{E}\cdot\boldsymbol{p}_{ij}\right)\left(Q^+_{ij}\right)^2$ and $\boldsymbol{a}_{12}=\boldsymbol{a},\,\boldsymbol{a}_{31}=\boldsymbol{b},\boldsymbol{a}_{32}=\boldsymbol{c}$.

\section{\label{appendixC} Effect of anisotropic Dzyaloshinskii-Moriya interaction}
 \begin{figure}[htb]
\includegraphics[width=0.85\textwidth]{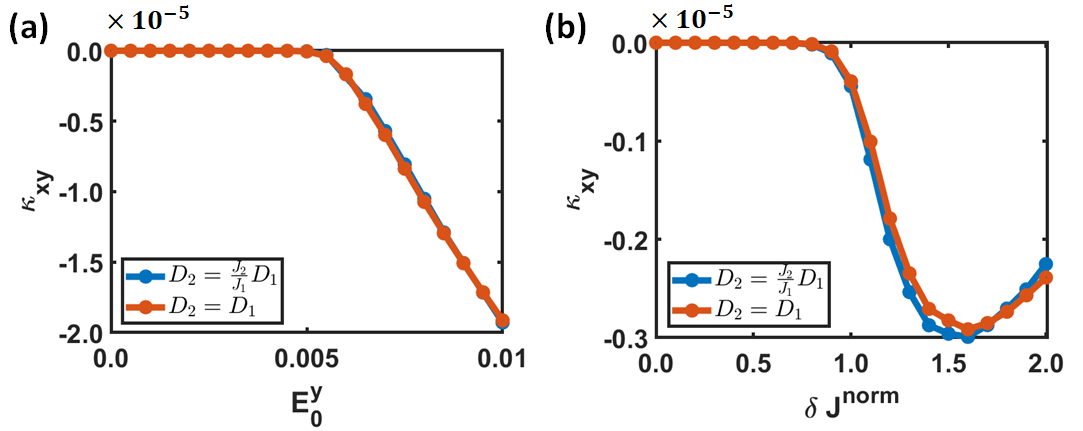} % this command will be ignored
\caption{ The thermal Hall conductance as a function of (a) electric field amplitude $E_0^y$ and (b) normalized breathing anisotropy $\delta J^{\text{norm}}=\delta J/\delta J^c$. The parameters of the system and electromagnetic field, if it is not varied, are $D_1=0.1J_1$, $\delta J=0.1J_1$, $p_0=1.0J_1$, $E_0^x=0.0$, $E_0^y=0.005$, $T=0.1J_1$, $\Omega=2.7J_1S$, $\gamma=10^{-3}$, $\eta=10^{-3}$. The system size for the simulations is $3\times 5000 \times 5000$, where $5000$ denotes the number of unit cells along each translation vector.}
\label{fig::AppendixFig}
\end{figure}

Our main text focuses on spin-exchange models on a breathing Kagome lattice, in which we only consider breathing anisotropy Heisenberg exchange interactions, leaving DMI invariant for simplicity.
In this appendix, we show that the effect of anisotropic DMI has very little effect on the qualitative as well as quantitative results.
The model Hamiltonian is given by,
\begin{align}
    \pazocal{H}_0=&-J_1\sum_{\left\langle i,j\right\rangle_1} \hat{\boldsymbol{ S}}_i\cdot\hat{\boldsymbol{S}}_j
    -J_2\sum_{\left\langle i,j\right\rangle_2} \hat{\boldsymbol{S}}_i\cdot\hat{\boldsymbol{S}}_j
    +D_1\sum_{\left\langle i,j\right\rangle_1}
    \nu_{ij} \hat{z}\cdot\left(\hat{\boldsymbol{S}}_i\times \hat{\boldsymbol{S}}_j\right)
    \nonumber\\
    &+D_2\sum_{\left\langle i,j\right\rangle_2}
    \nu_{ij} \hat{z}\cdot\left(\hat{\boldsymbol{S}}_i\times \hat{\boldsymbol{S}}_j\right)
    -B^z\sum_i \hat{S}_i^z.
    \label{eq::HamiltonianAppendix}
\end{align}
While the interactions $J_1=1.0$ and $D_1=0.1$ are kept fixed, the interactions $J_2=J_1+\delta J$ and $D_2=\frac{J_2}{J_1} D_1$ are varied by tuning the single parameter $\delta J$ which denotes breathing anisotropy.
Figures Fig.\,\ref{fig::AppendixFig}(a) and Fig.\,\ref{fig::AppendixFig}(b) show the thermal Hall conductivity for isotropic DMI\,(red curve) and anisotropic DMI\,(blue curve). According to the figures, the change in thermal Hall conductivity is very little when breathing anisotropy of DMI is taken into account.

\end{widetext}

\bibliographystyle{apsrev4-1}
%\bibliography{ref}
%merlin.mbs apsrev4-1.bst 2010-07-25 4.21a (PWD, AO, DPC) hacked
%Control: key (0)
%Control: author (72) initials jnrlst
%Control: editor formatted (1) identically to author
%Control: production of article title (-1) disabled
%Control: page (0) single
%Control: year (1) truncated
%Control: production of eprint (0) enabled
%

\end{document}